# Geometrical Eigen-subspace Framework Based Molecular Conformation Representation for Efficient Structure Recognition and Comparison

*Xiao-Tian Li[1], Xiao-bao Yang[1,2], and Yu-Jun Zhao[1,2]\**

[1]*Department of Physics and School of Materials Science and Engineering, South China University of Technology, Guangzhou, Guangdong 510640, China*
[2]*Key Laboratory of Advanced Energy Storage Materials of Guangdong Province, South China University of Technology, Guangzhou, Guangdong 510640, China*

[\*]Corresponding author. Tel: +86-20-87110426; fax: +86-20-87112837;
E-mail: zhaoyj@scut.edu.cn.

**Abstract:** We have developed an extended distance matrix approach to study the molecular geometric configuration through spectral decomposition. It is shown that the positions of all atoms in the eigen-space can be specified precisely by their eigen-coordinates, while the refined atomic eigen-subspace projection array adopted in our approach is demonstrated to be a competent invariant in structure comparison. Furthermore, a visual eigen-subspace projection function (EPF) is derived to characterize the surrounding configuration of an atom naturally. A complete set of atomic EPFs constitute an intrinsic representation of molecular conformation, based on which the interatomic EPF distance and intermolecular EPF distance can be reasonably defined. Exemplified with a few cases, the intermolecular EPF distance shows exceptional rationality and efficiency in structure recognition and comparison.

Molecular conformation plays an important role in solid state physics and material science for its intrinsic connection with various physical properties. A reasonable configuration space in which the vectors characterize structures uniquely provides us not only an insight of the intrinsic structures, but also a novel approach to material design, in particular in the age of Big Data. Recently, a flurry of global optimization methods were proposed for structure prediction [1-7], where a technique to eliminate equivalent configurations from a huge number of candidates is of great demand for efficiency. Meanwhile, the high efficiency is also imperative in chemical/structural similarity searching [8,9], which has been a basic retrieval mechanism in structure database. It is thus highly desirable to have a proper representation of molecular structures, based on which the intermolecular similarity can be reasonably defined and obtained. In chemistry and biology, while most of the research objects are valence-dominated organics, a traditional structural formula or a simplified notation such as SMILES [10,11] is sufficient to describe the molecular structure accurately. Moreover, by summarizing multiple atoms or functional groups into several types of nodes, one can get simpler molecular representations, such as pharmacophoric pattern [12-14] and reduced molecular graph [15]. The description and comparison of macromolecules based on these representations turn out to be easy to carry out. In solid state physics, however, it is hardly to find common structural features in most nanoparticles [16-20], where an unbiased description of molecular structure (robust enough against noise) and a related scheme to quantify the difference between configurations, are certainly helpful in the research community.

A complete set of atomic Cartesian coordinates can describe the detailed structure of a molecule accurately, but it is not convenient in structure comparison for its dependence on the coordinate frame and atomic ordering. It turns out that, a reasonable intermolecular distance based on the atomic coordination description needs to consider the placement of molecules and the matching of their atoms, which is hardly to carry out within polynomial time [21]. A similar issue is involved with distance matrix, another complete description of molecular structure. Despite the independence of coordinate frame, distance matrix also relies on the atomic ordering, which is imperative in structure comparison. Call *et al.* [1] proposed a coarse approach to arrange the atoms according to the atomic numbers and the distances to the center of mass of the molecule. The arrangement, however, is ambiguous and hardly to adopt in practice, since the atomic azimuths have not been taken into account.

Impeded by the problems in structure comparison, most of the structure prediction methods [3,6,7] prefer fingerprints/descriptors to characterize configurations rather than complete representations. Constructed by structural invariants, fingerprints/descriptors are convenient in structure comparison for the independence of coordinate frame and atomic ordering. A lot of molecular fingerprints/descriptors [3,7,22] were proposed based on interatomic distances, the vital invariants under translation, reflection and rotation in physics. Besides, the eigenvalues of distance matrix are also very suitable for structure recognition [21] and analysis [23,24]. Despite the advantages in structure comparison, fingerprints/descriptors encounter a problem that, they are hardly to guarantee a unique configuration with incomplete structural information. For instance, there may be two distinct molecules with identical eigenvalues, an isomorphism problem [25]. As a result, a high-qualified representation designed for structure recognition and comparison needs to balance accuracy and efficiency, i.e., it had better contain as much structural information as possible on the premise that the intermolecular distance can be easily carried out.

In this letter, we analyze the atomic projection information in the eigen-space based on an extended distance matrix, whose diagonal elements are assigned with the atomic numbers (or any other characters) of corresponding atoms. The derived eigen-subspace projection array (EPA) and eigen-subspace projection function (EPF) can characterize the surrounding configuration of an atom naturally, leading to the definition of the interatomic EPF distance and intermolecular EPF distance, which quantify the differences of atoms and molecules in geometric structure respectively. With discussions of several cases, the intermolecular EPF distance shows an excellent performance in structure recognition and comparison.

We define the elements of extended distance matrix of a molecule (or cluster) as

$$D_{ij} = z_i, \text{ if } i = j,$$
$$\quad\quad = d_{ij}, \text{ if } i \neq j, \tag{1}$$

where $z_i$ is the atomic number (or any other character) of atom $i$, and $d_{ij}$ is the Cartesian distance between atoms $i$ and $j$ in arbitrary units. The matrix differs from the conventional defined one in the diagonal elements by replacing zeros with the atomic numbers, crucial to distinguish elements. Such a matrix contains all the structural information of a molecule except overall chirality [26].

The extended distance matrix can be spectral factorized into a canonical form that

$$\mathbf{D} = \sum_{k=1}^{n} \lambda_k \mathbf{u}_k \mathbf{u}_k^{\mathrm{T}}, \tag{2}$$

where $\lambda_1, \lambda_2, \cdots, \lambda_n$ are all the eigenvalues of $\mathbf{D}$ in ascending order and $\mathbf{u}_k$ are the corresponding eigenvectors. The $i$-th component of eigenvector $\mathbf{u}_k$ represents the projection of atom $i$ on it. Typically, we can utilize the projections of an atom on the complete set of eigenvectors to characterize it in the eigen-space, i.e., a complete set of projections constitute the atomic eigen-coordinates with respect to the eigenvector basis. It is obvious that, the eigenvector basis (associated with the eigenvalues) together with the atomic eigen-coordinates, reserve all the information of the extended distance matrix.

Unfortunately, the eigenvector basis of identical molecule is not unique, especially in the degenerate cases that, one can arbitrarily choose mutually orthogonal eigenvectors associated to the degenerate eigenvalues. It turns out that the atomic eigen-coordinates vary with the eigenvector basis, which discourages their application in structure recognition and comparison. To overcome the randomness of eigenvector basis, we suggest replacing it with a framework constructed by a complete set of eigen-subspaces. Accordingly, the atomic projections on eigenvectors are replaced by the norm of projections on eigen-subspaces. Here we define the eigen-subspace projection array (EPA) of atom $i$ as

$$\mathbf{s}_i = \left( s_i^{\lambda_1}, s_i^{\lambda_2}, \cdots, s_i^{\lambda_s} \right), \tag{3}$$

where $\lambda_1, \lambda_2, \cdots, \lambda_s$ are all the distinct eigenvalues of $\mathbf{D}$ in ascending order, and $s_i^{\lambda_k}$ is the norm of orthogonal projection of atom $i$ on the eigen-subspace associated to $\lambda_k$. The eigen-subspace framework (associated with the distinct eigenvalues) together with the atomic EPAs constitute a general representation of molecular conformation.

Admittedly, the eigen-subspace framework and atomic EPAs overlook some structural details with respect to the eigenvector basis and atomic eigen-coordinates. In fact, all the molecular projection information on the eigen-subspace associated to $\lambda_k$ is contained in the projection matrix

$$\mathbf{P}^{\lambda_k} = \sum_m \mathbf{u}_m^{\lambda_k} \mathbf{u}_m^{\lambda_k \, \mathrm{T}}, \tag{4}$$

where $\mathbf{u}_m^{\lambda_k}$ constitute a complete set of eigenvectors associated to $\lambda_k$. It turns out that, the element $P_{ij}^{\lambda_k}$ of $\mathbf{P}^{\lambda_k}$ represents the dot product of projections of atoms $i$ and $j$, i.e., the diagonal elements reserve the individual projection information of the atoms, while the non-diagonal ones for their correlations. From this perspective, the EPAs discard the tangled correlations between atomic projections, crucial in structure recognition and comparison. With refined information, the EPAs become easy and efficient to operate, providing an intrinsic insight of atomic geometric positions.

For instance, Fig. 1(a) shows the detailed structure and corresponding extended distance matrix of methane, from which we can get the atomic eigen-coordinates and EPAs. Due to the symmetry of methane, there are a triply degenerate eigenvalues $\lambda_1 = \lambda_2 = \lambda_3 = -0.78$, associated to a three-dimensional eigen-subspace. The atomic eigen-coordinates specify the precise positions of all atoms in the eigen-space (especially in the triply degenerate eigen-subspace), but they rely on the eigenvector basis. In contrast, the atomic EPAs are very effective in structure recognition and comparison for their independence of eigenvector basis, although they can only specify the individual positions of atoms in the eigen-space (i.e., ignoring their correlation). Besides, equivalent atoms in a molecule would have identical EPAs naturally (such as the four equivalent H atoms in methane), critical in structure analysis.

It is worth noting that, the atomic EPA is not just a series of numbers, but associates with the eigen-subspaces closely. Suppose we stretch one of the C-H bonds of methane from 1.09 Å to 1.14 Å, as shown in Fig. 1(b). As the symmetry broken, the triple eigenvalues split into a single one and a double one, leading to the splitting of the atomic EPAs. Note that the deviated H atom has a quite different EPA compared with the other three, although they have similar positions in the molecule. Besides, it is hardly to determine the structural similarity for the ideal and slightly variant methane molecules by their atomic EPAs directly (Fig. 1). These irregularities would be easy to understand while the eigen-subspaces are taken into account. In fact, since $\sum_{k=1}^{s}\left(s^{\lambda_k}\right)^2=1$, each atom can be viewed as a unit vector in the eigen-space, whose segments are projected on the complete set of eigen-subspaces. From this perspective, the deviated H atom in the variant methane has a segment of $0.86^2$ projected on the eigen-subspace associated to $\lambda_1=-0.84$, while the other three H atoms have segments of $0.29^2$ and $0.82^2$ on $\lambda_1=-0.84$ and $\lambda_2=\lambda_3=-0.78$ correspondingly, as seen in Fig. 1(b). Since similar segments of atomic vectors are projected on the split eigen-subspaces associated to numerically similar eigenvalues, the structural similarity between the H atoms in variant methane can be perceived.

To determine the structural similarity directly and more conveniently, one can get assisted from the atomic eigen-subspace projection function (EPF), which illustrates the relationship between the segments of a unit atomic vector in the eigen-space ($S\in[0,1]$) and the eigenvalues that the projected eigen-subspaces are associated to. The atomic EPFs of the variant methane are shown in Fig. 1(b), from which we easily observe that the red line (corresponding to the deviated H atom) is no remarkable different to the green ones (the other three H atoms), while the blue line (the C atom) is significantly different, consistent with the atomic positions in the molecule. Note that all EPFs increase monotonously since we have sorted the eigenvalues in ascending order beforehand. As the EPF summarizes the surrounding configuration of an atom finely, a complete set of atomic EPFs constitute an intrinsic representation of molecular conformation.

Based on the atomic EPF, we define the EPF distance between atoms $i$ and $j$ to be

$$d_{i,j}^{\text{EPF}} = \int_0^1 |\Lambda_i - \Lambda_j| dS, \tag{5}$$

where $\Lambda_i$ and $\Lambda_j$ represent the EPFs of atoms $i$ and $j$ respectively. Note that the EPF distance quantifies the atomic difference in their surrounding configurations, different from their geometric distance in real space. For instance, in the variant methane, $d_{\text{HH}}^{\text{EPF}} = 0$, while $d_{\text{H'H}}^{\text{EPF}} = 0.08$ (Fig. 1(b), where H' stands for the deviated H atom), revealing the equivalence of the three fixed H atoms and the degree of deviation of the stretched one.

Moreover, we point out here that, the atomic EPF depends on the atomic number (or any other atomic character assigned in the diagonal element of $\mathbf{D}$) to a great extent besides its surrounding configuration, since $\int_0^1 \Lambda_i dS = z_i$. As a result,

$$d_{i,j}^{\text{EPF}} = \int_0^1 |\Lambda_i - \Lambda_j| dS \geq \left| \int_0^1 \Lambda_i dS - \int_0^1 \Lambda_j dS \right| = |z_i - z_j|. \tag{6}$$

It turns out that the EPF distance between atoms of different elements is always greater than the difference between their atomic numbers, critical to distinguish elements. For instance, the C atom in the variant methane has an exact EPF distance of 5 to all the H atoms as seen in Fig. 1(b), since its EPF is always larger than the ones of H atoms.

Following the interatomic EPF distance, we define the EPF distance between molecules $p$ and $q$ (both with $n$ atoms) to be

$$d_{p,q}^{\text{EPF}} = \frac{1}{n} \min_{\{i,j\}} \sum_{i=j=1}^n d_{i,j}^{\text{EPF}}, \tag{7}$$

where atoms $i$ and $j$ belong to molecules $p$ and $q$ respectively, and the summation represents taking the accumulated difference of all atoms. Here each atom $i \in p$ is associated to an atom $j \in q$, and we take the atoms of two molecules matched in such a way that the accumulated distance is minimized. This is actually an assignment problem, which can be easily approached by the Hungarian algorithm [27].

As mentioned above, the eigenvalues of distance matrix encounter the isomorphism problem in structure recognition that, there may be two distinct configurations with identical eigenvalues. For instance, Fig. 2(a) shows two twenty-one-atom triangular fragments (TFs) with identical eigenvalues despite their structural nonequivalence, clearly reflected by the shadowed outlines. While the eigenvalues are incapable of distinguishing the two TFs, we can get assisted from the atomic projection information on the eigen-subspaces. Here we derive the atomic eigen-spectra from the EPAs for a better view by Lorentz expansion

$$I_i(\lambda) = \sum_{k=1}^{s} s_i^{\lambda_k} \frac{\sigma/\pi}{(\lambda - \lambda_k)^2 + \sigma^2}, \qquad (8)$$

where $\lambda_1, \lambda_2, \cdots, \lambda_s$ are all the distinct eigenvalues of $\mathbf{D}$ in ascending order, $s_i^{\lambda_k}$ is the EPA of atom $i$ on the eigen-subspace associated to $\lambda_k$, and $\sigma$ is the Lorentzian width parameter (set to be 0.2 here). In Fig. 2, the atoms of the two TFs are associated with each other according to their interatomic EPF distances in ascending order by the Hungarian algorithm. Although the atomic eigen-spectra of the two TFs have peaks at the same positions in general due to their identical eigenvalues, the structural nonequivalence can be clearly indicated, especially from the inset of their difference (Fig. 2(b)). Their intermolecular EPF distance corresponds to be 0.331 (see Eq. (7)). Of note, the spectra of atoms 1, 2, and 3 of the two TFs are identical since they have the same distances with other 20 atoms in both TFs, reflecting a reasonable association between the atoms in the molecules. To further demonstrate the efficient and rational approach of EPF for the association of atoms between the molecules, we present an example of $B_{36}$ clusters in the supplemental material.

While the surrounding configuration of each atom can be depicted by an EPF, molecules turn out to locate in a configuration space constructed by a series of EPFs. Such a configuration space provides us not only a visual knowledge of the geometric correlations between configurations, but also some specific approaches to practical issues. For instance, we explore the potential energy surface of $LJ_{38}$ cluster (LJ cluster is consisted of identical atoms interacting by a pair Lennard-Jones potential) and project 200 local minima on a 2D map (Fig. 3), from which we can get an intuitive impression of the configurations. In particular, the global minimum $LJ_{38}$-$O_h$ is far from the other local minima in contrast to the second-minimum $LJ_{38}$-$C_{5v}$, illustrating clearly why $LJ_{38}$-$O_h$ is very difficult to obtain in most global optimization methods. Besides, we can conjecture the potential transition paths between configurations with the knowledge of their geometric correlations, critical to determine the transition states.

Based on the extended distance matrix, a representation has been proposed to describe the molecular conformation by projecting the atomic geometric information on the eigen-subspaces. Neglecting the correlations between atomic projections, the refined EPA and EPF reflect the surrounding configuration of an atom finely. Moreover, the EPF based approach is demonstrated to be *competent* in the isomorphism problem, and very *intuitive* and *efficient* in structure comparison.

This work is financially supported by NSFC (Grant 11574088, 51431001), the Foundation for Innovative Research Groups of the National Natural Science Foundation of China (Grant No.: 51621001), and Natural Science Foundation of Guangdong Province of China (Grant No. 2016A030312011).

FIG. 1. The detailed structures, extended distance matrices, atomic eigen-coordinates, EPAs and EPFs of (a) the methane and (b) the variant methane with one of its C-H bonds stretched from 1.09 Å to 1.14 Å. Here, the notations for equations ① and ② are explained in the main text. In equation ③, the summation runs over the complete set of eigenvectors associated to eigenvalue $\lambda_k$. For instance, the first column of EPA in (a) is simply the square root of sum of the squares of the first three columns of the eigen-coordinates correspondingly.

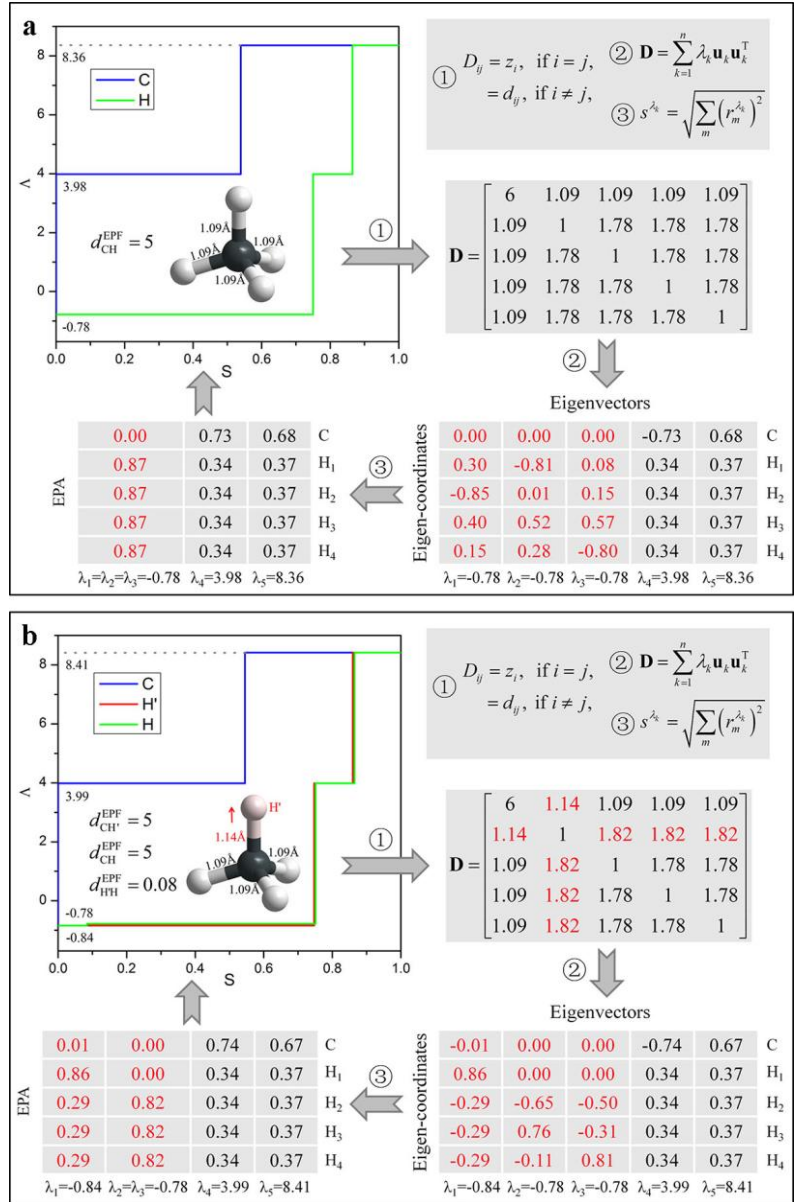

FIG. 2. (a) Two TFs with distance matrices of identical eigenvalues and (b) their atomic eigen-spectra with an inset of their difference. The atoms of the two TFs are matched and arranged according to their interatomic EPF distances, which are marked on the right side of the eigen-spectra. Note that the interatomic EPF distances are consistent with the differences of the atomic eigen-spectra, indicating their rationality.

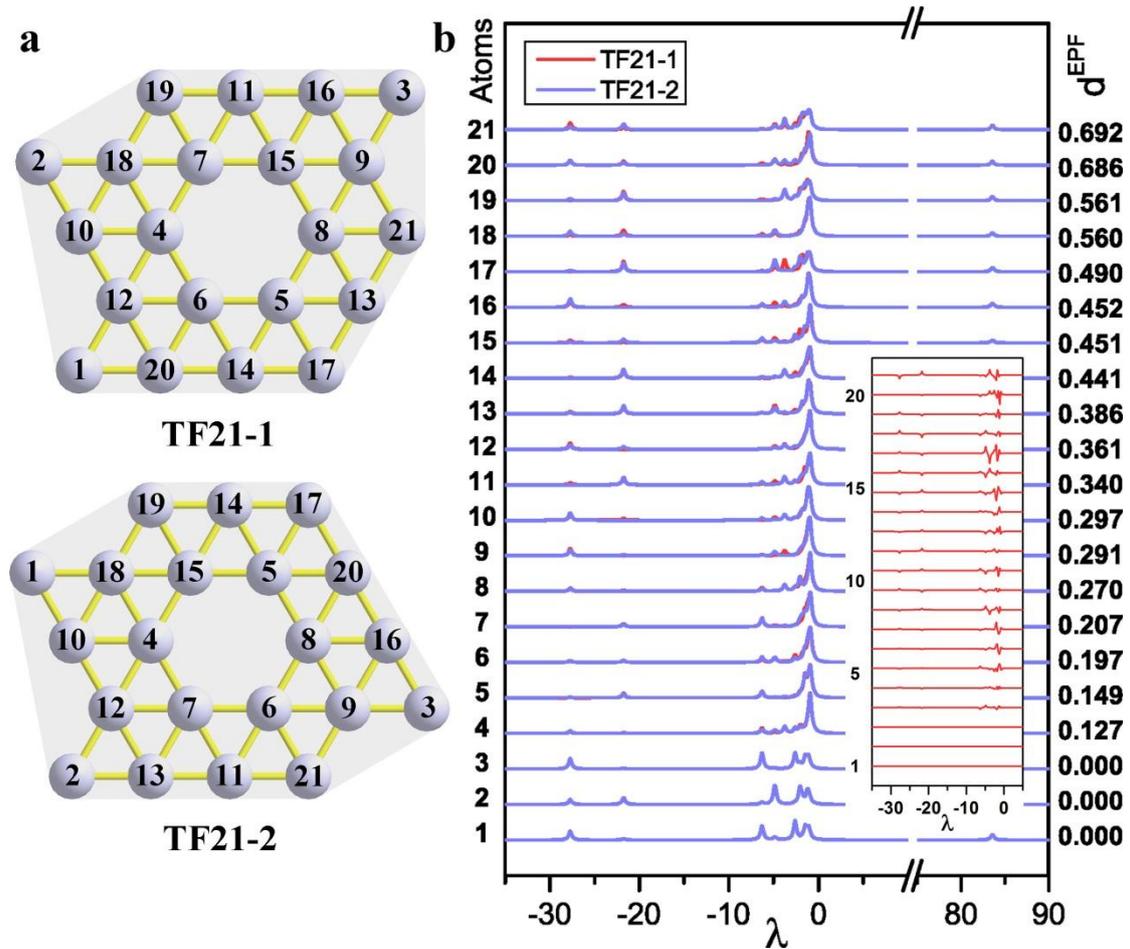

FIG. 3. 2D landscape of the potential energy surface of $LJ_{38}$ cluster. Each point here represents a local minimum configuration, whose position on the 2D map is determined by minimizing $\sum_{p<q}\left(d_{pq}^{2D}-d_{pq}^{EPF}\right)^2$. The energies of the configurations are indicated by their colors, while the global minimum $LJ_{38}$-$O_h$ and the second-minimum $LJ_{38}$-$C_{5v}$ are marked by pentagram particularly.

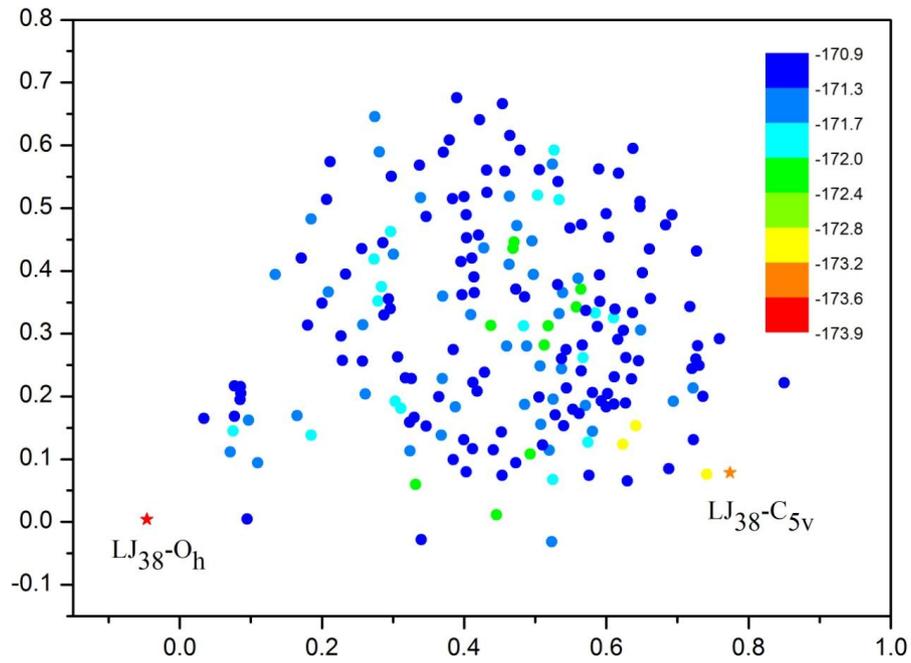

**S1: Rationality of intermolecular EPF distance**

As mentioned in the main text, a reasonable intermolecular distance based on the atomic Cartesian coordination description needs to consider the placement of molecules and the matching of their atoms. For instance, Sadeghi *et al.* [1] proposed a distance between molecules $p$ and $q$ by

$$\text{RMSD}(p,q) = \frac{1}{\sqrt{n}} \min_{\mathbf{P},\mathbf{U}} \left\| \mathbf{R}^p - \mathbf{U}\mathbf{R}^q\mathbf{P} \right\|,$$

where $\mathbf{R} \equiv (r_1, r_2, \cdots, r_n) \in \mathbb{R}^{3 \times n}$ is a detailed atomic coordination description of a molecule, the absolute symbol represents taking the Frobenius norm for coordinates of $p$ and $q$, while $\mathbf{U}$ and $\mathbf{P}$ stand for a rotation (and reflection) and a permutation operation respectively. Since the effect of translation has been eliminated beforehand by taking the centroid of each molecule to be the origin, the distance quantifies the molecular structural difference intrinsically, irrespective of the coordinate frame and the atomic ordering. Unfortunately, there is no algorithm nowadays to find the global RMSD within polynomial time.

In order to test the accuracy and efficiency of the intermolecular EPF distance, we compare it with the RMSD distance. We perform a molecular dynamics (MD) simulation at 300K for the ground state of $B_{36}$ clusters [2,3], a quasiplanar configuration with a central hexagonal hole. The MD simulation is carried out based on the density functional theory (DFT) implemented in the Vienna *ab initio* simulation package (VASP) [4,5]. The projector augmented wave (PAW) and the Perdew-Burke-Ernzerhof (PBE) of Generalized Gradient

Approximation (GGA) functional [6] are employed for the calculations. The cutoff energy is 500 eV and the vacuum distance is set to be 20 Å. Using Nosé-thermostat [7,8] approach, we performed the constant-temperature MD simulation for $B_{36}$ with a time step of 1 femtosecond.

The RMSD and EPF distances between each pair of 50 configurations, taken from MD simulation after equilibrium, are shown in Fig. S1. It is obvious that there is a strong positive correlation between the RMSD and EPF distances with a correlation coefficient of 0.986. Of note, the RMSD distances, difficult to carry out in general, are obtained here for $B_{36}$ clusters benefited from their quasiplanar structures. We rotate (and reverse/reflect) one of the two given configurations along the normal of the quasiplane for every $\frac{2\pi}{500}$ radian, and perform the permutation and the rotation (and reflection) operation alternately until the RMSD distance reaches a local minimum. Since the two configurations have been matched considering a lot of potential azimuths, we can get the global minimum from all the local minima. However, this is a time-consuming process, expending about 500 times as much as that by the EPF distance approach. Moreover, the RMSD approach is hardly to generalize to three dimensional configurations, while our EPF distance has no such a limit. It turns out that, our intermolecular EPF distance shows a good consistency with the RMSD distance with much higher efficiency.

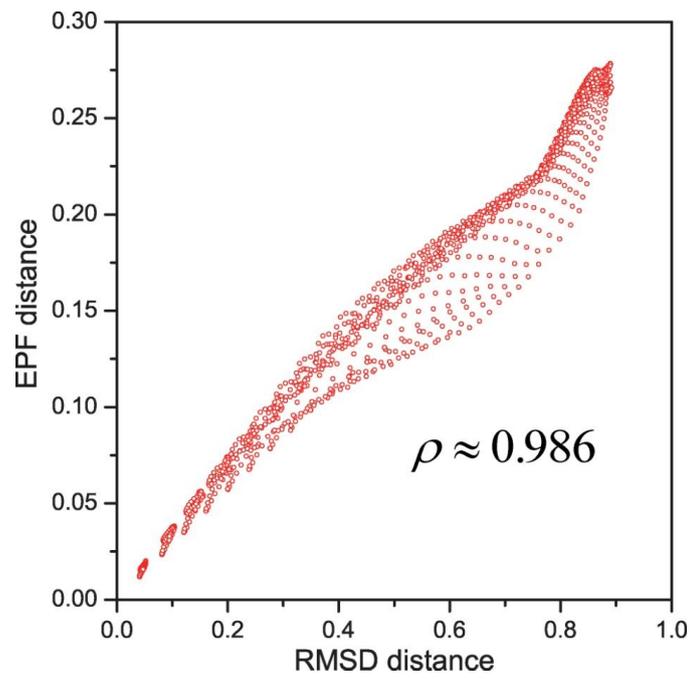

FIG. S1. Correlation of RMSD and EPF distances between various pairs of 50 $B_{36}$ configurations taken from the MD simulation.